\documentclass[number,preprint,3p]{elsarticle}
\usepackage{newpxtext}

\usepackage[T1]{fontenc}

\usepackage{threeparttable}
\usepackage{geometry}
\usepackage{soul}
\usepackage{ulem}
\usepackage{float}
\usepackage{xcolor}
\usepackage[framemethod=TikZ]{mdframed}

\usepackage{amsmath, hyperref, geometry, mathrsfs, amssymb, amsthm}
\usepackage{mathrsfs}
\hypersetup{colorlinks = true}
\usepackage{booktabs}
\usepackage{multirow}
\usepackage{footnote}
\usepackage{mathtools}
\usepackage{enumitem}
\usepackage{nomencl}
\usepackage{changepage}
\makesavenoteenv{tabular}
\usepackage{textcomp}
\usepackage{cleveref}
\usepackage{subcaption}
\usepackage{color}
\usepackage{tikz}
\usepackage{dsfont}
\usepackage{algorithm}
\usepackage{algpseudocode}
\usepackage[titletoc,toc,title]{appendix}
\crefrangeformat{equation}{Eqs.~(#3#1#4) --~(#5#2#6)}

\usepackage[labelfont=bf]{caption}
\graphicspath{{./picture/}}
\usepackage[perpage]{footmisc}

\def\r{\mathbb{R}}

\def\tr{^\intercal}
\newcommand{\Prob}[1]{\mathbb{P}\left(#1\right)}
\newcommand{\E}[1]{\mathbb{E}\left[#1\right]}

\newcommand{\vect}[1]{\boldsymbol{#1}}

\makenomenclature
\usepackage{ifthen}
    \renewcommand{\nomgroup}[1]%
        {\ifthenelse{\equal{#1}{S}}{\item[\textbf{Subscripts}]}{}}

\usepackage{lineno}
\usepackage{makecell}
\usepackage[export]{adjustbox}

\soulregister\cite7
\soulregister\citep7
\soulregister\citet7
\soulregister\ref7
\soulregister\eqref7
\soulregister\pageref7

\journal{arXiv}

\begin{document}

\begin{frontmatter}
\renewcommand{\thefootnote}{\fnsymbol{footnote}}

\title{Maximum entropy-based modeling of community-level \\hazard responses for civil infrastructures}

\author{Xiaolei Chu}
\ead{xiaoleichu@berkeley.edu}
\author{Ziqi Wang\texorpdfstring{\corref{correspondingauthor}}{}}
\ead{ziqiwang@berkeley.edu}
\address{Department of Civil and Environmental Engineering, University of California, Berkeley, CA, 94720, United States}

\cortext[correspondingauthor]{Corresponding author: Ziqi Wang}

\begin{abstract}
Perturbed by natural hazards, community-level infrastructure networks operate like many-body systems, with behaviors emerging from coupling individual component dynamics with group correlations and interactions. It follows that we can borrow methods from statistical physics to study the response of infrastructure systems to natural disasters. This study aims to construct a joint probability distribution model to describe the post-hazard state of infrastructure networks and propose an efficient surrogate model of the joint distribution for large-scale systems. Specifically, we present maximum entropy modeling of the regional impact of natural hazards on civil infrastructures. Provided with the current state of knowledge, the principle of maximum entropy yields the ``most unbiased'' joint distribution model for the performances of infrastructures. In the general form, the model can handle multivariate performance states and higher-order correlations. In a particular yet typical scenario of binary performance state variables with knowledge of their mean and pairwise correlation, the joint distribution reduces to the Ising model in statistical physics. In this context, we propose using a dichotomized Gaussian model as an efficient surrogate for the maximum entropy model, facilitating the application to large systems. Using the proposed method, we investigate the seismic collective behavior of a large-scale road network (with 8,694 nodes and 26,964 links) in San Francisco, showcasing the non-trivial collective behaviors of infrastructure systems. 

\end{abstract}
\end{frontmatter}

\renewcommand{\thefootnote}{\fnsymbol{footnote}}

\section*{Highlights}
\begin{itemize}
  \item Maximum entropy modeling for regional hazard responses of infrastructure systems.
  \item A surrogate of the maximum entropy model for large-scale systems.
  \item Collective seismic behaviors of a large-scale road network. 
\end{itemize}

\newpage


\section{Introduction} 
\label{sec:intro}
\noindent Civil structures and infrastructures, when forming into networked systems such as transmission (e.g., water, gas, and power) and transportation networks, serve as the backbone of a community. In assessing their risks from natural hazards, stakeholders are not only interested in the performance of individual structures but also the community-level integrity and functionality. Such assessments require accurate and efficient modeling of infrastructure network responses to hazards \cite{ouyang2014review}. However, the collective hazard behavior of infrastructures is intricate owing to a large number of components, complex network topology, component inter-dependencies, and incomplete knowledge. 

To reveal community-level regularities, the primary step is understanding the direct impacts of hazards on infrastructure systems. Due to the aleatory uncertainties in the occurrence and intensity of hazards, probabilistic approaches are dominant. In performance-based engineering, fragility functions are widely used to describe the structure's probability of exceeding limit states (such as performance/damage levels) as a function of hazard intensities. In performance-based earthquake engineering, Incremental Dynamic Analysis (IDA) \cite{vamvatsikos2002incremental,vamvatsikos2004applied}, Multiple Stripe Analysis (MSA) \cite{jalayer2003direct,baker2015efficient}, cloud analysis \cite{miano2018cloud}, and extended fragility analysis \cite{andriotis2018extended}, are widely recognized approaches for fragility analysis. It was shown in \cite{yia2022appraisal} that IDA provides an upper bound of the actual structural fragility and the cloud method only provides suboptimal fragility curves due to its inherent assumptions in the objective function; the MSA was shown to be a subcase of the Bernoulli model introduced by Shinozuka et al. \cite{shinozuka2000statistical}, which is then part of the extended fragility analysis framework. In performance-based wind engineering \cite{ciampoli2011performance}, fragility analysis emphasizes more on the limit states related to serviceability and comfort \cite{cui2018unified, larsen2016engineering,chu2021probabilistic}. Fragility functions are typically defined at the structure-level, i.e., a specific fragility curve is assigned to each of the civil structures in a region of interest. In this context, the correlation between the performance states of different structures is contributed by the interdependencies of local intensity measures of hazards; it is not straightforward to consider the contribution from the similarity of structures. With the ever-increasing computational power, a simulation-based approach using probabilistic hazard and physics-based infrastructure models becomes viable to complement fragility analysis \cite{jayaram2008statistical,park2007modeling,shaw2018physics,graves2011cybershake}. Physics-based simulations can capture complex interdependencies among structures, at the cost of computing power and interpretability. 

In this paper, we present maximum entropy modeling \cite{jaynes1957information} as a complementary perspective to fragility and simulation-based approaches. 
The maximum entropy modeling has gained much success in biostatistics \cite{shlens2006structure}, such as inferring direct interactions from gene expression data \cite{stein2015inferring}, exploring spatial sources of disease outbreaks \cite{ansari2022inferring}, and uncovering patterns in brain activity \cite{ashourvan2021pairwise}. The method has the potential to reveal interactions between different groups of network components. However, its applicability in regional hazard impact assessment for civil infrastructures has yet to be  investigated. In this study, we will show the procedures of building maximum entropy-based models for the responses of infrastructure systems to hazards. Furthermore, we emphasize the collective behavior in hazard responses of infrastructure components. This concept is rarely discussed in civil engineering, but it is often stressed in neuroscience \cite{shlens2006structure}, social sciences \cite{blumer1971social}, and material science \cite{ji2023collective}. Weak local, microscopic correlations can trigger nonlocal, macroscopic behaviors \cite{shlens2006structure}. In the context of civil engineering, a meaningful collective behavior is the system-level transition from functioning to failure. Using the maximum entropy modeling equipped with the lens of statistical physics \cite{landau2013statistical,castellano2009statistical}, we can investigate the phase transitions \cite{romanczuk2023phase} and critical phenomena to better understand the hazard resilience of infrastructures. Incidentally, it is worth mentioning that collective behaviors could trigger cascading behaviors \cite{guo2017critical}, but the former focuses on spontaneous responses while the latter emphasizes sequential propagation processes.

A substantial drawback of the maximum entropy model is the poor scalability toward larger systems. Learning a maximum entropy model is an important topic of Boltzmann machine learning \cite{ackley1985learning}. For high-dimensional models, a computational bottleneck in the training process is the sample estimates for the mean and cross-moment values of a high-dimensional random vector; this computation needs to be performed at each iterative learning step. Since independent sampling from the maximum entropy model is typically infeasible, the sample estimates are often obtained from a Markov Chain Monte Carlo (MCMC) algorithm \cite{fischer2012introduction,fischer2014training}, which is computationally demanding. Approximate approaches such as contrastive divergence learning \cite{hinton2002training} and pseudo maximum likelihood estimations \cite{besag1975statistical,besag1977efficiency} have been proposed to accelerate the training of maximum entropy models, at the cost of accuracy \cite{brook1964distinction} and stability \cite{du2020improved}. Moreover, an MCMC algorithm will be required again in sampling from the trained maximum entropy model. In this paper, we adopt an alternative route of using near-maximum entropy models. Specifically, we investigate the use of a dichotomized Gaussian model \cite{macke2009generating}, a near-maximum entropy model \cite{wohrer2019ising} allowing for independent sampling, as an efficient surrogate of the maximum entropy model for large-scale systems. 

This paper is organized as follows. Section \ref{Sec:MaxEnt} introduces the formulation of maximum entropy models for regional hazard responses of infrastructure systems. Section \ref{Sec:Para} presents the training process for the maximum entropy model. Section \ref{Sec:Surr} introduces a surrogate model for the maximum entropy model. Section \ref{Sec:App} demonstrates the proposed method in analyzing the collective behavior of a large-scale road network under an earthquake scenario. Section \ref{Sec:Addi} discusses possible future research topics and Section \ref{Sec:Concl} provides concluding remarks.

\section{Maximum entropy modeling}\label{Sec:MaxEnt}
\noindent The principle of maximum entropy states that, among all probability distributions that satisfy the current state of knowledge about a system, the distribution that maximizes the information entropy is the most unbiased, and thus the ``best", representation of the system \cite{jaynes1957information}. Entropy, in this context, is a measure of uncertainty. A distribution with maximum entropy is the least informative in the sense that it makes the least amount of assumptions beyond the known constraints. The maximum entropy model has significant potential in modeling the collective hazard responses of networked civil systems. In this section, we will propose a general model that can accommodate to various hazards.

\subsection{A brief introduction to the principle of maximum entropy}
\noindent Consider a random variable $X\in\{s_1, s_2, \dots,s_n\}$ representing the discrete performance state of a structure. In performance-based engineering, $s_i$ can represent performance levels such as ``immediate occupancy", ``life safety", ``collapse prevention", and ``collapse". Let $p(x)$ denote the unknown distribution of $X$, i.e., $p(x)\equiv\mathbb{P}(X=x)$. We seek the distribution $p(x)$ under the \textit{constraints}, i.e., current state of knowledge, of generalized moments expressed by
\begin{equation}\label{eq:Moment}
\langle f_i(X) \rangle = \sum_{x} p(x) f_i(x)\,,i=1,2,\dots,m\,,
\end{equation}
where $f_i$ are functions of $X$ and the expectations $\langle f_i(X) \rangle$ are known, e.g., collected from data, estimated from computational models, etc. It is worth mentioning that the form of Eq.~\eqref{eq:Moment} is fairly general to represent various quantities collected in practice. For example, on top of the conventional moments where $f_i(x)=x^k$, $k\in\mathbb{N}^+$, Eq.~\eqref{eq:Moment} can also represent probability constraints if $f_i(x)=\mathds{1}(x-s_k)$, where $\mathds{1}$ is a binary indicator function for $x-s_k=0$. 

The problem of determining $p(x)$ given Eq.~\eqref{eq:Moment} can be ill-defined if the constraints are insufficient for a unique solution. Naturally, information theory \cite{jaynes1957information} enters the picture as a theoretical foundation to introduce additional assumptions. Specifically, information theory defines the information entropy as a quantity/functional accounting for the ``amount of uncertainty'' encoded in a probability distribution:


\begin{equation}
H(X) = -\sum_{x} p(x) \log p(x)\,.
\end{equation}
Under the constraints of Eq.~\eqref{eq:Moment}, the most ``unbiased'' probability distribution must maximize the entropy. It follows that the Lagrangian multipliers $\lambda$ and $\mu_i$ can be introduced to maximize  the following equation
\begin{equation}
\begin{aligned}
  {L}(p) = &-\sum_{x} p(x) \log p(x) + \lambda \left(\sum_{x} p(x)-1\right) \\&+ \sum_{i}\mu_i \left(\sum_{x} p(x) f_i(x) - \langle f_i(X) \rangle\right)  \,.
\end{aligned}
\end{equation}
The solution takes the Boltzmann distribution form:
\begin{equation}\label{eq:GenSol}
p(x)=\frac{1}{Z}\exp\left(\sum_{i}\mu_i f_i(x)\right)\,,
\end{equation}
where $\mu_i$ and $Z$ can be determined from the constraints together with the normalization condition; the normalizing constant $Z$ is also called partition function in statistical physics.

The equations introduced above can be extended to the multivariate case such that $x$ is replaced by vector $\vect x$ representing the joint state of multiple structures. Hereafter, we provide a simple example for illustration purposes. Assume there are three individual structures in a network. Their joint state is denoted as $\vect{x}=(x_{1},x_{2},x_{3})\tr\in\{0,1\}^3$, where $0$ denotes ``safe" and $1$ ``failure". Our current state of knowledge is their individual failure probabilities, i.e., $\langle X_{i} \rangle$, $i=1,2,3$. It follows that the maximum entropy joint distribution is 
\begin{equation}
\label{eq:demo_maxent}
p(\vect{x})=\frac{1}{Z} \exp \left(\sum_{i=1}^{3} \mu_{i} x_{i} \right)\,,
\end{equation}
where $\mu_{i}$ are ``free" parameters of the model that need to be inferred from the constraints. Notice that $Z$ is not a free parameter because it is coupled with $\mu_{i}$ by the normalization condition. Since there is no second-order information on $\vect X$, the maximum entropy solution yields independence between the three random variables. The maximum entropy distribution  is a parsimonious model with the minimum number of free parameters that is consistent with the current state of knowledge. The number of free parameters is equal to the number of independent constraints.

\subsection{Modeling regional hazard response under cross-moment constraints}
\noindent For a community of $d$ structures, let $\vect X=(X_1,X_2,\dots,X_d)\tr\in\{s_1,s_2,\dots,s_n\}^d$ denote their (random) performance states. We assume that the constraints are \textit{cross-moments}, then the $f_i(\vect x)$ in Eq.~\eqref{eq:Moment} has the form:
\begin{equation}\label{eq:crossMom}
    f_i(\vect x)={x_1^{k^{i}_1}\cdot{x_2^{k^{i}_2}}\cdot...\cdot x_d^{k^{i}_d}}\,,\,i=1,2,...,m\,,
\end{equation}
where $k^{i}_1,k^{i}_2,...,k^{i}_d\in\mathbb{N}$. The previous example with  Eq.~\eqref{eq:demo_maxent} corresponds to cross-moment constraints of order 1, i.e., $\sum_j k^i_j=1$. If second or higher-order information is collected, the maximum entropy distribution takes the form
\begin{equation}
\label{eq:bi_maxent}
p(\vect{x})=\frac{1}{Z} \exp \left(\sum_i \mathcal{H}_i x_i+\sum_{i,j} \mathcal{J}_{i j} x_i x_j+\sum_{i,j,k} \mathcal{K}_{i j k} x_i x_j x_k+\cdots\right)\,,
\end{equation}
where to simplify the notations we let the summations run over $\{1,2,...,d\}$; this has to be accompanied by setting the parameter to zero if the corresponding cross-moment is not collected. For example, suppose $\E{X_1X_2}$ is not collected, we set $\mathcal{J}_{1,2}\equiv0$. The total number of free parameters in Eq.~\eqref{eq:bi_maxent} should equal the number of constraints, $m$. 

In civil engineering practice, collecting information regarding the first and second-order cross-moments are most common. In this context, if we further focus on a binary failure/safe performance state, Eq.~\eqref{eq:bi_maxent} drops the third and higher-order terms and reduces to the Ising model in statistical physics. On the other hand, if the number of performance levels approaches infinity, we can model $\vect X$ as continuous. In this case, the maximum entropy model under mean and covariance constraints is the multivariate Gaussian.

\subsection{The Ising model}
\noindent Let $\vect X=(X_1,X_2,\dots,X_d)\tr\in\{0,1\}^d$, and further assume that the first and second-order cross-moments are collected. The maximum entropy distribution admits a compact form:
\begin{equation}
\label{eq:ising2}
p(\vect{x};\vect{J})=\frac{1}{Z(\vect{J})}\exp\left(-\mathrm{H}(\vect x;\vect{J})\right)=\frac{1}{Z(\vect{J})}\exp \left(\vect{x}\tr \vect{J} \vect{x}\right)\,,
\end{equation}
where $\vect{J}$ is a $d\times d$ matrix, and it is seen that the \textit{Hamiltonian} $\mathrm{H}$ is quadratic. This compact form is derived using the identity $x_i^2=x_i$ for $x_i\in\{0,1\}$, indicating the first-order terms can be absorbed into the diagonal entries of $\vect J$. The conventional Ising model assumes a binary state of $\{-1,1\}$ and thus requires an explicit decomposition of first and second-order terms (see Eq.\eqref{eq:bi_maxent}). However, this difference is trivial because the number of free parameters stays the same, i.e., equals the number of independent constraints, regardless of the superficial equation form. Suppose the full covariance matrix\footnote{Here, if the full covariance matrix of $\vect X$ is known, it would be redundant to specify the mean vector.} of $\vect X$ is collected, $\vect{J}$ has $(d^2+d)/2$ parameters/entries to be determined. This task is challenging for a large $d$.

\section{Parameter Identification}\label{Sec:Para}
\noindent In this section, we focus on identifying the Ising model parameters, while the algorithms discussed here can be applied to general maximum entropy models.

\subsection{Preliminary formulations}
\noindent For applications of the Ising model to regional hazard responses, we may meet two scenarios: (1) random samples of the performance states of structures, i.e., samples of $\vect X$, are collected, and (2) the first and second-order cross-moments of $\vect X$ are specified. The first scenario becomes relevant if a regional-scale physics-based simulator is available, so that it can generate random structural responses under a stochastic hazard model. The second scenario fits the situation where one has empirical models for failure probabilities and pairwise correlations, e.g., fragility functions and spatial correlation models for hazard responses. It is worth mentioning that there exists alternative scenarios, such as a ``one-shot" sample of regional hazard response is collected from post-hazard survey. This ``third" scenario needs to be augmented by additional models/assumptions/data to make the inference possible, such that it will eventually be transformed into scenario (1) or (2). In this work, we address the parameter identification for both scenarios. 


We start by assuming that $N$ independent and identically distributed observations $\mathcal{D} = \left\{\vect{x}^{(i)}\right\}_{i=1}^N$ on $\vect X$ are collected, where $\vect{x}^{(i)} \in \{0, 1\}^{d}$; this assumption can be relaxed in a later stage. We need to find the parameter matrix $\vect{J}$ that maximizes the likelihood function
\begin{equation}
\begin{aligned}
\vect J^*&=\arg\max_{\vect J}\mathcal{L}\left(\vect{J}\right)\\
&=\arg\max_{\vect J}\prod_{\vect x\in\mathcal{D}}p(\vect{x};\vect{J})\,,    
\end{aligned}
\end{equation}
where $\mathcal{L}$ is the likelihood function and $p(\vect{x};\vect{J})$ is from Eq.~\eqref{eq:ising2}. If we  have some prior knowledge on $\vect J$, i.e., specifying a prior distribution $p(\vect{J})$, we can extend the point estimation of $\vect J$ into a posterior distribution $p(\vect{J}|\mathcal{D}) \propto \mathcal{L}\left(\vect{J}\right)p(\vect J)$. In this work, we focus on the more basic question of point estimation by maximum likelihood, leaving the Bayesian parameter estimation to future studies. 

The log-likelihood function is
\begin{equation}
\begin{aligned}
\ell(\vect J)=&\ln \mathcal{L}(\vect{J}) = \sum_{\vect x\in\mathcal{D}} \left(-\mathrm{H}(\vect{x};\vect{J}) - \ln Z(\vect{J}) \right) \\
& = \sum_{\vect{x}\in \mathcal{D}} -\mathrm{H}(\vect{x};\vect{J}) - N\ln \left(\sum_{\vect{x}\in\Omega_{\vect X}} \exp \left(-\mathrm{H}(\vect{x};\vect{J})\right) \right)\,,
\end{aligned}
\end{equation}
where $\Omega_{\vect X}$ denotes the sample space of $\vect X$, containing $2^d$ elements. The negative log-likelihood can be viewed as an energy function with state variables $\vect J$, and we seek the ground state such that the energy is minimized. Clearly, the computational bottleneck of the log-likelihood function for large $d$ is the evaluation of the second summation term, which is infeasible to be computed exactly. 

The gradient of $\ell(\vect{J})$ can be written as
\begin{equation}
\label{eq:gradient_comp}
\begin{aligned}
\frac{\partial \ell(\vect{J})}{\partial \vect{J}} & = -\sum_{\vect x\in\mathcal{D}} \frac{\partial \mathrm{H}(\vect{x};\vect{J})}{\partial \vect{J}} + N \sum_{\vect{x} \in\Omega_{\vect X}} \frac{\exp \left(-\mathrm{H}(\vect{x};\vect{J})\right)}{Z(\vect{J})} \frac{\partial \mathrm{H}(\vect{x};\vect{J})}{\partial \vect{J}} \\ & = -N\left(\frac{1}{N}\sum_{\vect{x}\in \mathcal{D}} \frac{\partial \mathrm{H}(\vect{x};\vect{J})}{\partial \vect{J}} - \sum_{\vect{x} \in \Omega_{\vect X}} \frac{\exp \left(-\mathrm{H}(\vect{x};\vect{J})\right)}{Z(\vect{J})} \frac{\partial \mathrm{H}(\vect{x};\vect{J})}{\partial \vect{J}}\right) \\ & = -N\left\langle \frac{\partial \mathrm{H}(\vect{x};\vect{J})}{\partial \vect{J}} \right\rangle_{\text{data}} +  N\left\langle\frac{\partial \mathrm{H}(\vect{x};\vect{J})}{\partial \vect{J}} \right\rangle_{\text{model}}\,.
\end{aligned}
\end{equation}
The last line of Eq.~\eqref{eq:gradient_comp} is interpretable: the first ``expectation" \footnote{More precisely, it is a sample mean approximation for the expectation with respect to the unknown distribution that generates the data.} is taken with respect to data, while the second expectation is with respect to the parametric model for a specified $\vect J$. Since the maximum likelihood estimation for $\vect J$ seeks Eq.~\eqref{eq:gradient_comp} being zero, it follows that the ``data" and ``model" terms should be equal. This is intuitively expected: if the expectation estimated from data matches that of the distribution model, the data is indeed generated by the model and the corresponding $\vect J$ is ``correct". Recall that $\mathrm{H}(\vect{x};\vect{J})=-\vect{x}\tr\vect{J}\vect{x}$, the gradient $\partial \ell (\vect{J}) / \partial \vect{J}$ has the explicit form: 
\begin{equation}
\label{eq:gradient_ell}
\begin{aligned}
\frac{\partial \ell(\vect{J})}{\partial \vect{J}} & = N\langle \vect{x}\vect{x}\tr \rangle_{\text{data}} -N \langle \vect{x}\vect{x}\tr \rangle_{\text{model}}  \\
& = \sum_{\vect{x} \in \mathcal{D}} \vect{x}\vect{x}\tr - \frac{N}{Z(\vect{J})}\sum_{\vect{x}\in\Omega_{\vect X}}\vect{x}\vect{x}\tr\cdot\exp\left(-\vect{x}\tr\vect{J}\vect{x}\right)\,.
\end{aligned}
\end{equation}
Since the maximum likelihood solution seeks $\frac{\partial \ell(\vect{J})}{\partial \vect{J}}=0$, the equation for $\vect J$ is 
\begin{equation}\label{eq:solveJ}
    \frac{1}{Z(\vect{J})}\sum_{\vect{x}\in\Omega_{\vect X}}\vect{x}\vect{x}\tr\cdot\exp\left(-\vect{x}\tr\vect{J}\vect{x}\right)=\frac{1}{N}\sum_{\vect{x} \in \mathcal{D}} \vect{x}\vect{x}\tr\,.
\end{equation}
This equation suggests that the aforementioned two scenarios regarding the form of collected data can be treated in a unified manner. Specifically, if samples of $\vect X$ are collected, the summation in the right-hand side of Eq.~\eqref{eq:solveJ} can be evaluated using the data. If the mean and cross-correlation of $\vect X$ are specified (from empirical fragility and correlation models), the right-hand side of Eq.~\eqref{eq:solveJ} can be regarded as an expectation that is directly given.

\subsection{Gradient descent solutions}
\noindent Identifying parameters of the Ising model is an important topic of Boltzmann machine learning, where the Ising model is also known as the fully visible Boltzmann machine (VBM) \cite{ackley1985learning}. We can use the gradient descent algorithm to find the approximate solution of Eq.~\eqref{eq:solveJ}:
\begin{equation}
\label{eq:ML_grad_desc}
\vect{J}^{(\tau + 1)} = \vect{J}^{(\tau)} - \vect{\eta} \circ \left(\langle \vect{x}\vect{x}\tr \rangle_{\text{data}} - \langle \vect{x}\vect{x}\tr \rangle_{p(\vect{x};\vect{J}^{(\tau)})}\right)\,,
\end{equation}
where $\vect{\eta}$ is the learning rate/step size, which can be adaptively tuned depending on the specific algorithmic implementation; $\circ$ means the entry-wise product. This updating equation can be extended to the general maximum entropy model described by Eq.~\eqref{eq:bi_maxent}, such that:
\begin{subequations}
\begin{align}
\mathcal{H}_{i}^{(\tau + 1)} & = \mathcal{H}_{i}^{(\tau)} - \eta_{\mathcal{H}} \left(\langle x_{i} \rangle_{\text{data}} - \langle x_{i} \rangle_{p(\vect{x};\mathcal{H}^{(\tau)},\mathcal{J}^{(\tau)},\mathcal{K}^{(\tau)},\dots)} \right) \,,\\
\mathcal{J}_{ij}^{(\tau + 1)} & = \mathcal{J}_{ij}^{(\tau)} - \eta_{\mathcal{J}} \left(\langle x_{i}x_{j} \rangle_{\text{data}} - \langle x_{i}x_{j} \rangle_{p(\vect{x};\mathcal{H}^{(\tau)},\mathcal{J}^{(\tau)},\mathcal{K}^{(\tau)},\dots)} \right)\,, \\
\mathcal{K}_{ijk}^{(\tau + 1)} & = \mathcal{K}_{ijk}^{(\tau)} - \eta_{\mathcal{K}} \left(\langle x_{i}x_{j}x_{k} \rangle_{\text{data}} - \langle x_{i}x_{j}x_{k} \rangle_{p(\vect{x};\mathcal{H}^{(\tau)},\mathcal{J}^{(\tau)},\mathcal{K}^{(\tau)},\dots)} \right)\,,
\end{align}
\end{subequations}
where $\eta_{\mathcal{H}}$, $\eta_{\mathcal{J}}$, and $\eta_{\mathcal{K}}$ are learning rates. Following the pattern, one can write the updating equations for the fourth and higher-order terms. 

In Eq.~\eqref{eq:ML_grad_desc}, as expected, the computational bottleneck is the evaluation of the expectation $\langle \vect{x}\vect{x}\tr \rangle_{p(\vect{x};\vect{J}^{(\tau)})}$. For high-dimensional problems with large $d$, it is infeasible to exhaust the $2^{d}$ combinations. One solution is to use Monte Carlo simulation, i.e., using Markov Chain Monte Carlo techniques \cite{desjardins2010tempered} to generate random samples from $p(\vect{x};\vect{J}^{(\tau)})$ and approximating $\langle \vect{x}\vect{x}\tr \rangle_{p(\vect{x};\vect{J}^{(\tau)})}$ by samples. However, this process can be time consuming because the MCMC sampling may require a long chain to reach equilibrium. An alternative is to use the Contrastive Divergence (CD) learning \cite{hinton2002training,carreira2005contrastive}. The idea is to run MCMC for a few steps (typically only one step) and use those premature samples to approximate the expectation $\langle \vect{x}\vect{x}\tr\rangle_{p(\vect{x};\vect{J}^{(\tau)})}$. The theoretical argument is to replace the gradient of the original Kullback-Leibler (KL) divergence between data and model\footnote{Notice that maximizing the likelihood of observing the data is equivalent to minimizing the KL divergence between the data and the model, because the marginal likelihood $p(\mathcal{D})$ is a constant independent of $p(\vect x;\vect J)$.} by that of a Contrastive Divergence, expressed by \cite{hinton2002training}
\begin{equation}\label{eq:CD}
\mathrm{CD}_{n} = \mathrm{KL}(p_{0} \| p_{\infty}) - \mathrm{KL}(p_{n} \| p_{\infty})\,,
\end{equation}
where $p_{0}$ denotes the distribution of the data, $p_{n}$ denotes the distribution associated with running the Markov chain for $n$ steps, and $p_{\infty}$ denotes the equilibrium distribution, i.e., the model. It is seen that $p_{\infty}$ is cancelled out in the definition of CD, thus the gradient of Eq.~\eqref{eq:CD} does not involve evaluating $p_{\infty}$. It follows that the updating equation of $\vect{J}$ in CD learning is
\begin{equation}
\label{eq:CD_grad_desc}
\vect{J}^{(\tau + 1)} = \vect{J}^{(\tau)} - \vect{\eta} \circ \left(\langle \vect{x}\vect{x}\tr \rangle_{\text{data}} - \langle \vect{x}\vect{x}\tr \rangle_{p_{n}} \right)\,.
\end{equation}
Theoretically speaking, CD learning is a biased algorithm; however, empirical results suggest the bias is typically small \cite{carreira2005contrastive}. It is worth mentioning that there are alternative approaches to CD learning, such as score matching \cite{hyvarinen2005estimation,hyvarinen2007some}, maximum pseudolikelihood learning \cite{besag1975statistical}, minimum velocity learning \cite{movellan1993learning}, and minimum probability flow learning \cite{sohl2011new}. In practice, we found that CD learning works better for cases with weak pairwise correlation, where the estimation of the marginal distribution is more accurate. Deep learning methods can be potentially useful to accelerate the training of maximum entropy models. A recent study \cite{noe2019boltzmann} showed that the deep learning method can be leveraged to efficiently sample from maximum entropy models even when the modes are well-separated.


\subsection{Mean-field approximations}
\noindent Although Boltzmann learning with gradient descent algorithms can lead to accurate approximations of the maximum entropy model parameters, it is usually slow. There are many approximate solutions based on the mean-field theory, such as naive mean-field theory, independent-pair approximation, Sessak-Monasson approximation, and inversion of TAP equation \cite{roudi2009ising}. These approximate methods can be used to generate warm starting points for optimization algorithms, but their accuracy depends largely on the network size and correlation structure \cite{roudi2009statistical}.





\subsection{An illustrative example}
\label{sec:40_dim_example}
\begin{figure}[H]
    \centering
    \makebox[\textwidth][c]{\includegraphics[]{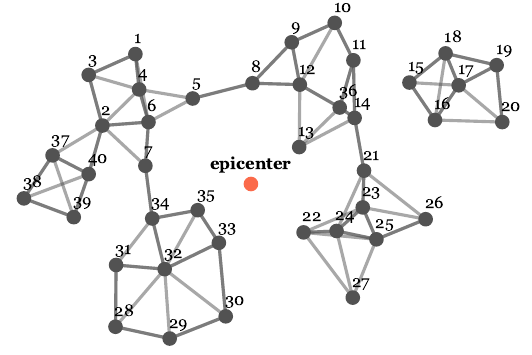}}
    \caption{\textbf{A simple road network with 40 nodes}. \textit{The network topology, i.e., how the nodes are connected, is irrelevant at the stage of modeling the nodal performances; but it will become relevant when studying the global functionality of the network.}}
    \label{fig:road_network_10d}
\end{figure}

\begin{figure}[H]
    \centering
    \makebox[\textwidth][c]{\includegraphics[scale=0.67]{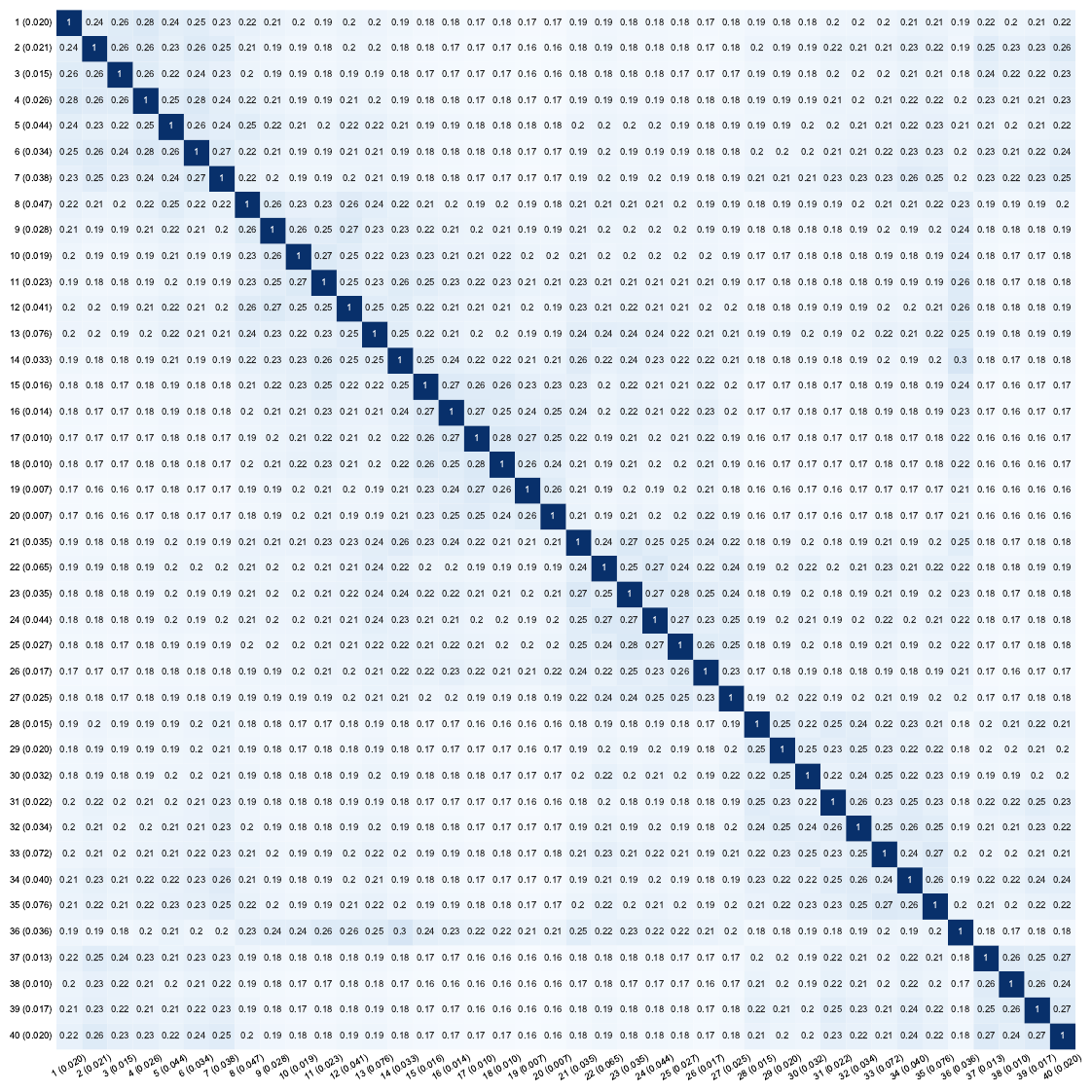}}
    \caption{\textbf{Failure probabilities and their cross-correlations}.  \textit{The axis labels represent the node numbers with the failure probability of that node in brackets; the diagonal entries are $1$; the off-diagonal entries are the Pearson correlation coefficients.}}
    \label{fig:corr_mu_map}
\end{figure}

\noindent In this section, we use a simple example to illustrate the maximum entropy modeling.  We consider a network with 40 nodes subjected to the impact of hazards (Figure~\ref{fig:road_network_10d}); each node has a binary state of $\{0\text{(work)},1\text{(fail)}\}$. The current knowledge/constraint is assumed to be the mean value and correlation matrix of the state vector, {which is created by the method in Sec.~\ref{sec:seismic_risk_assessment}}. Therefore, the maximum entropy model is the Ising model expressed by Eq.~\eqref{eq:ising2}. Fig.~\ref{fig:corr_mu_map} shows the failure probabilities of nodes and their correlation coefficients. 

\begin{figure}[H]
    \centering
    \begin{adjustbox}{minipage=\linewidth,scale=0.9}
    \makebox[\textwidth][c]{
    \begin{subfigure}[c]{90mm}
      \centering
      \includegraphics[]{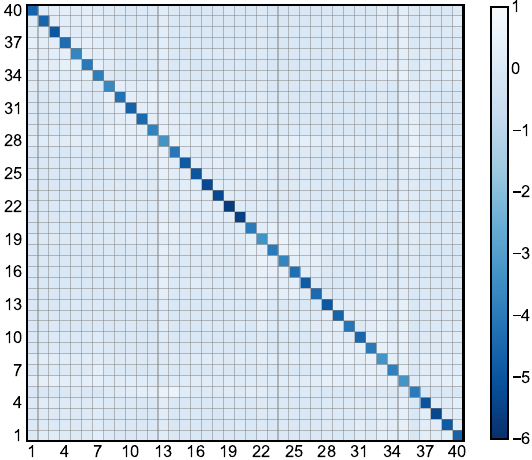}
      \caption{Identified $\vect{J}$ of the Ising model}
      \label{fig:J_mat}
    \end{subfigure}
    \begin{subfigure}[c]{90mm}
      \centering
      \includegraphics[]{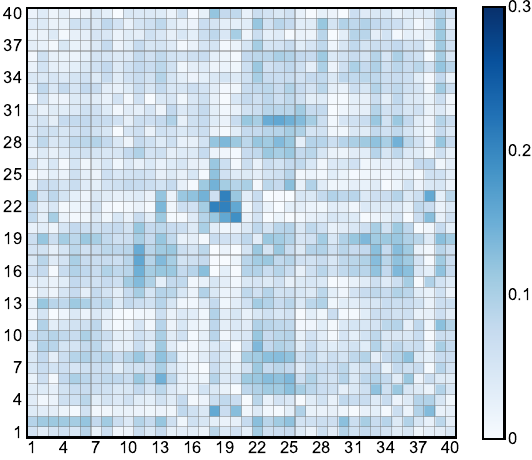}
      \caption{Identification error of the covariance matrix}
      \label{fig:error_mat}
    \end{subfigure}
    }
    \end{adjustbox}
    \caption{\textbf{Identified $\vect{J}$ and its error for the 40-dimensional Ising model}. \textit{Given a positive correlation matrix, the entries of $\vect J$ can take negative values. The errors in the covariance matrix are typically less than $4\%$. The errors are expected to be lower with a more carefully designed learning process, e.g., adaptive learning rate or more Gibbs samples at each iteration}.}
\end{figure}

The gradient descent algorithm expressed by Eq.~\eqref{eq:ML_grad_desc} is used to estimate $\vect J$. The starting point $\vect J^{(0)}$ is initialized randomly, and $\langle \vect{x}\vect{x}\tr \rangle_{p(\vect{x};\vect{J}^{(\tau)})}$ is estimated by Gibbs sampling using $100,000$ samples with a burn-in period of $20,000$ samples. The gradient descent algorithm is iterated for $2,000$ steps, and the learning rate is set to $\eta=0.2$. Fig.~\ref{fig:J_mat} shows the identified $\vect{J}$, and Fig.~\ref{fig:error_mat} shows the errors in the covariance matrix reconstructed from the trained model. The trained maximum entropy model can be leveraged to investigate the collective behavior and global functionality of the network. A concrete example will be presented in a later section. Finally, the parameters of the Ising model have clear statistical physics interpretations, suggesting future research directions of bringing theories of phase transitions and critical phenomena to the regional analysis of civil engineering systems.

\section{Dichotomized Gaussian distribution as a surrogate for the Ising model}\label{Sec:Surr}
\noindent For large-scale infrastructure networks with numerous components, the high-dimensional Boltzmann learning becomes computationally infeasible, and it is meaningful to seek efficient approximations for maximum entropy models. The Dichotomized Gaussian distribution \cite{macke2009generating,emrich1991method} is an attractive candidate for a surrogate of the Ising model. This model has been successfully applied to investigate the pairwise correlations and collective behaviors in neural populations \cite{macke2009generating}. 

\subsection{The dichotomized Gaussian model}
\noindent The main idea of the dichotomized Gaussian approximation is to treat the binary state vector as a filtered continuous Gaussian vector. The filtering is deterministic and fixed; therefore, the model will be uniquely determined if the mean and covariance matrix of the latent Gaussian vector are specified. Specifically, assume that the binary random state variables $\vect{X}\in\{0,1\}^d$ are generated from latent Gaussian variables $\vect{Z}\in\r^d$, such that
\begin{equation}\label{eq:dicho}
\vect{X} = \mathds{1}_{\geq0}(\vect{Z}),\ \vect{Z} \sim \mathcal{N}(\vect{\gamma},\vect{\Lambda})\,,
\end{equation}
where $\mathds{1}_{\geq0}$ is a vector indicator function that maps each component $Z_i\geq0$ to $1$ and $Z_i<0$ to $0$. Assigning unit variances for the latent Gaussian distribution \cite{macke2009generating,emrich1991method}, i.e., $\Lambda_{ii}=1$, $i=1,2,...,d$, the mean $\boldsymbol{\mu}$ and covariance matrix $\vect{\Sigma}$ of $\vect{X}$ can be expressed as
\begin{equation}\label{eq:paraest}
\begin{aligned}
\mu_{i} &= \Phi(\gamma_{i})\,,\\ 
\Sigma_{ii} & = \Phi(\gamma_{i})\left(1 - \Phi\left(\gamma_{i}\right)\right) = \Phi(\gamma_{i})\Phi(-\gamma_{i})\\ 
\Sigma_{ij} & =\Psi(\gamma_{i},\gamma_{j};\Lambda_{ij})\,, \ \mathrm{for} \ i\neq j\,,
\end{aligned}
\end{equation}
where $\Psi(x,y;\lambda) = \Phi(x,y;\lambda) - \Phi(x)\Phi(y)$; $\Phi(\cdot)$ is the cumulative distribution function (CDF) for the standard univariate Gaussian distribution, and $\Phi(\cdot,\cdot;\lambda)$ denotes the joint CDF of the bivariate Gaussian with unit variances and a pairwise correlation $\lambda$. The parameters $\vect\mu$ and $\vect\Sigma$ are observable, while $\vect\gamma$ and $\vect\Lambda$ are latent. Therefore, we need to use Eq.~\eqref{eq:paraest} to find $\vect\gamma$ and $\vect\Lambda$ given $\vect\mu$ and $\vect\Sigma$. To determine $\vect\gamma$, we simply use the inverse function $\gamma_{i} = \Phi^{-1}(\mu_{i})$. To determine $\Lambda_{ij}$, we need to numerically solve the one-dimensional equation $\Psi(\gamma_{i},\gamma_{j};\Lambda_{ij}) = \Sigma_{ij} $. This is straightforward because the function $\Psi(\gamma_{i},\gamma_{j};\Lambda_{ij})$ is monotonic in $\Lambda_{ij}$.

Provided with $(\vect\gamma,\vect\Gamma)$, the joint probability mass function of the dichotomized Gaussian distribution is
\begin{equation}
q(\vect{x})=\frac{1}{(2 \pi)^{d / 2}|\vect{\Lambda}|^{1 / 2}} \int_{a(x_1)}^{b(x_1)} \cdots \int_{a(x_d)}^{b(x_d)} \exp \left(-\frac{1}{2}(\vect{z}-\vect{\gamma})\tr \vect{\Lambda}^{-1}(\vect{z}-\vect{\gamma})\right)\,\mathrm{d}\vect{z}\,,
\end{equation}
where $a_{i}=0$, $b_{i}=\infty$ if $x_{i}=1$; $a_{i}=-\infty$, $b_{i}=0$ if $x_{i}=0$. This PMF expression is listed here for completeness. In fact, the PMF expression is seldom used in most applications, because for large systems, a generic performance metric $\langle g(\vect{X}) \rangle$ is typically  approximated by Monte Carlo simulation. The random samples of $q(\vect x)$ can be easily generated via Eq.~\eqref{eq:dicho}, without resorting to MCMC algorithms. To summarize, the computational advantage of the dichotomized Gaussian distribution over the Ising model lies both in parameter estimation and random sampling; the cost is being a near maximum entropy model. 


\subsection{Numerical verification in covariance matrix}
\begin{figure}[H]
    \centering
    \makebox[\textwidth][c]{\includegraphics[]{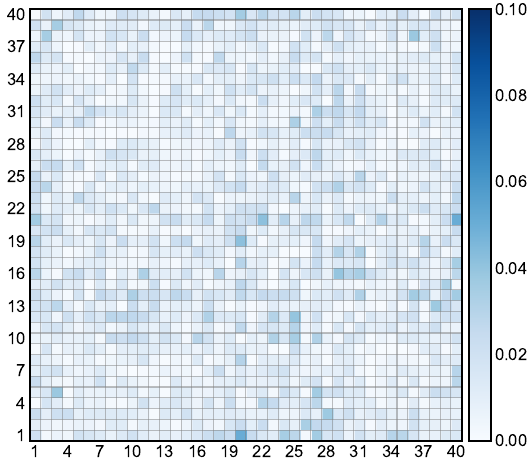}}
    \caption{\textbf{Error of the covariance matrix using samples generated by dichotomized Gaussian distribution.} \textit{It is observed that the identification error is usually less than $0.4\%$.}}
    \label{fig:error_DG}
\end{figure}

\noindent We reproduce the example in Sec.~\ref{sec:40_dim_example} using the dichotomized Gaussian model and compute the covariance matrix for a comparison. As shown in Fig.~\ref{fig:error_DG}, the identification error in the covariance matrix is negligible; this result is expected because the mapping between the covariance matrices of the latent Gaussian and observable binary vectors are relatively simple. The training time for the dichotomized Gaussian model is around 1.5 seconds, while the Boltzmann learning for the Ising model takes around $5.5$ minutes. 

\subsection{Numerical verification in entropy}
\noindent Since the Ising model maximizes the entropy given the first- and second-order cross-moment constraints, we need to compare the entropy between the Ising and dichotomized Gaussian models. We let the number of components, or size, of the system vary from $2$ to $40$ and compare the entropy obtained from the Ising and dichotomized Gaussian models. The technical details on estimating the entropy using variance reduction methods are introduced in \ref{sec:appendixA}. The failure probabilities and their pairwise correlations at different sizes are adopted from Fig.~\ref{fig:corr_mu_map}, such that the system with size $j$ consists of nodes $\{1,2,...,j\}$ in Fig.~\ref{fig:corr_mu_map}.

As shown in Fig.~\ref{fig:entro_corr}, the entropies of the two models are close. Theoretically speaking, the entropy of the Ising model should always be larger; this condition can be breached in practice due to parameter identification and sampling variabilities. The entropy difference is expected to be small if the correlation is weak. For a simple illustration, we consider the extreme case of no correlation. The Ising model reduces to
\begin{equation}
    p(\vect{x}) = \frac{1}{Z} \exp{\left(\sum_{i=1}^{d}\mathcal{H}_{i}x_{i}\right)} = \frac{1}{Z}\prod_{i=1}^{d} \exp{\left(\mathcal{H}_{i}x_{i}\right)} = \prod_{i=1}^{d}p_i(x_{i})\,,
\end{equation}
where $p_i(1)$ is the failure probability of node $i$. The dichotomized Gaussian model reduces to
\begin{equation}
\begin{aligned}
q(\vect{x}) & =\frac{1}{(2 \pi)^{d / 2}|\vect{I}|^{1 / 2}} \int_{a(x_1)}^{b(x_1)} \cdots \int_{a(x_d)}^{b(x_d)} \exp \left(-\frac{1}{2}(\vect{z}-\vect{\gamma})\tr \vect{I}^{-1}(\vect{z}-\vect{\gamma})\right)\,\mathrm{d}\vect{z} \\
& = \prod_{i=1}^{d} \frac{1}{\sqrt{2\pi}} \int_{a(x_{i})}^{b(x_{i})} \exp{\left(-\frac{1}{2}\left( z_{i} - \gamma_{i} \right)^{2}\right)} \mathrm{d}z_{i} \\
& = \prod_{i=1}^{d} p_{i}(x_{i})\,.
\end{aligned}
\end{equation}

{In this context, $p(\vect{x}) \equiv q(\vect{x})$ are the independent 
 multivariate Bernoulli distribution, and the entropy achieves the maximum (see Fig.~\ref{fig:entropy_noncorr}), i.e., for any joint distribution $p(x_1,x_2,...,x_d)$, we must have
\begin{equation}
\begin{aligned}
    H\left( X_{1},X_2,\dots,X_{d} \right) & = -\sum_{x_{1}\in \{ 0,1 \}} \sum_{x_{2}\in \{ 0,1 \}} \cdots \sum_{x_{d}\in \{ 0,1 \}} p\left( x_{1},x_2,\dots,x_{d} \right) \log  p\left( x_{1},x_2,\dots,x_{d} \right) \\
    & \leq - \sum_{i=1}^{d} p_{i}(x_{i}) \log p_{i}(x_{i})\,.
\end{aligned}
\end{equation}
\begin{figure}[H]
    \centering
    \begin{adjustbox}{minipage=\linewidth,scale=0.9}
    \makebox[\textwidth][c]{
    \begin{subfigure}[c]{90mm}
      \centering
      \includegraphics[]{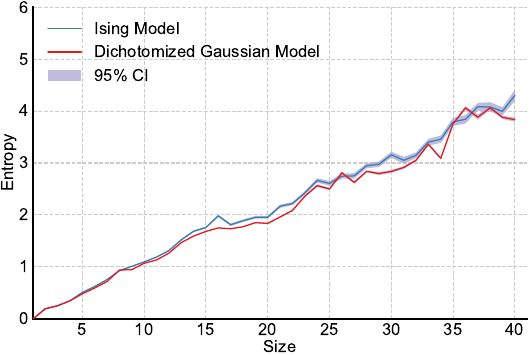}
      \caption{\textbf{With pairwise correlation extracted from Fig.}~\ref{fig:corr_mu_map}}
      \label{fig:entro_corr}
    \end{subfigure}
    \begin{subfigure}[c]{90mm}
      \centering
      \includegraphics[]{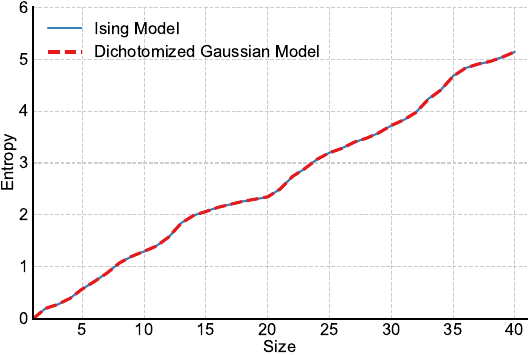}
      \caption{\textbf{Without correlation.} \textit{Nodes are independent}}
      \label{fig:entropy_noncorr}
    \end{subfigure}
    }
    \end{adjustbox}
    \caption{\textbf{Comparison of the entropy between the Ising and dichotomized Gaussian models}. \textit{The entropy difference between the Ising and dichotomized Gaussian models in our tested example is small. The right panel is theoretically trivial, but it verifies our algorithmic implementations. }}
    \label{fig:entropy_comparison}
\end{figure}


\section{Application: Seismic collective behaviors of the road network in San Francisco}\label{Sec:App}
\label{sec:application_SF}
\noindent In this section, we will illustrate an important application of the maximum entropy modeling–-the assessment of global functionalities and collective behaviors of infrastructure systems. We analyze the post-earthquake behaviors of road networks in San Francisco. For illustrative purposes, an empirical seismic hazard model\cite{lee2021multi} is adopted to generate the mean values of road failures and their pairwise correlations. Since this road network has $8,694$ nodes and $26,964$ links, a dichotomized Gaussian distribution is established as a surrogate for the underlying maximum entropy model, i.e., the Ising model. The global functionalities and collective behaviors of the road network are analyzed using random samples generated from the near-maximum entropy model. 
\subsection{Seismic hazard model}
\label{sec:seismic_risk_assessment}
\subsubsection*{First-order constraint}
\noindent We adopt the seismic hazard model developed in \cite{goda2008spatial,lee2021multi} to generate the first and second-order cross-moments to build the near maximum entropy model. Specifically, assuming a log-normal fragility curve for each of the road components, the mean of $X_i$ is expressed by
\begin{equation}
\E{X_i}=\Prob{X_i=1}=\Phi\left(\frac{\bar D_i-\bar C_i}{\sqrt{\sigma_{D_i}^2+\sigma_{C_i}^2}}\right)\,,
\end{equation}
where $\Phi(\cdot)$ is the cumulative distribution function (CDF) of the standard Gaussian distribution; $\bar D_i$ is the average seismic demand in terms of peak ground acceleration (PGA) and $\bar C_i$ average seismic capacity; $\sigma_{D_i}^{2}$ and $\sigma_{C_i}^{2}$ are variances of the random demand and capacity, respectively. To model the average seismic demand $\bar D_i$, we adopt the empirical attenuation relation in \cite{lim2012efficient}, expressed by
\begin{equation}
\bar D_i=-0.5265+\left(-0.3303+0.0599\left(M_w-4.5\right)\right) \ln \left(r_i^2+1.35^2\right)-0.0115 \sqrt{r_i^2+1.35^2}\,,
\end{equation}
where $M_{w}$ is the earthquake magnitude; $r_{i}$ is the distance between the epicenter and the $i$th road component in kilometers. Adopting the parameter values of \cite{lee2021multi}, we assume $\sigma_{D_i}^2=0.32$, $\sigma_{C_i}^2=0.48$, and $\bar C_i=\ln{(0.85)}$ for all components. 
\subsubsection*{Second-order constraint}
\noindent The correlation coefficient between $X_i$ and $X_j$ is modeled by \cite{lee2021multi}
\begin{equation}
\label{eq:correlation_network}
\rho_{X_iX_j}=\frac{\sigma_D^2 \rho_{D_i D_j}+\sigma_{C_i} \sigma_{C_j} \delta_{i j}}{\sqrt{\sigma_D^2+\sigma_{C_i}^2} \sqrt{\sigma_D^2+\sigma_{C_j}^2}}\,,
\end{equation}
where $\delta_{ij}$ is the Kronecker delta, and $\rho_{D_i D_j}$ represents the correlation coefficient between the PGAs of two sites, expressed by \cite{goda2008spatial,lee2021multi}
\begin{equation}
\rho_{D_i D_j}=\frac{\sigma_\eta^2}{\sigma_{D}^2}+\rho_{\varepsilon_i \varepsilon_j}\left(\Delta_{i j}\right) \frac{\sigma_{\varepsilon}^2}{\sigma_{D}^2}\,,
\end{equation}
where $\sigma_\eta^2$ is the variance of the random inter-event residual, $\sigma_{\varepsilon}^2$ is the variance of the random intra-event residual, and $\sigma_{D}^2=\sigma_\eta^2+\sigma_{\varepsilon}^2$; we set $\sigma_\eta^2=0.07$ and $\sigma_{\varepsilon}^2=0.25$. Finally, $\rho_{\varepsilon_i \varepsilon_j}\left(\Delta_{i j}\right)$ is modeled by the intra-event spatial correlation model \cite{lim2012efficient}, expressed as
\begin{equation}
\rho_{\epsilon_{i} \epsilon_{j}} (\Delta_{ij}) = \exp (-0.27\Delta_{ij}^{0.40})\,,
\end{equation}
where $\Delta_{ij}$ is the distance between site $i$ and $j$.

Fig.~\ref{fig:community_network} shows the marginal failure probabilities of the road segments in San Francisco. The epicenter is chosen as $37.80^{\circ}N, 122.27^{\circ}W$. The $26,964\times26,964$ correlation matrix is obtained from Eq.~\eqref{eq:correlation_network}.
\begin{figure}[H]
    \centering
    \begin{adjustbox}{minipage=\linewidth,scale=0.9}
    \makebox[\textwidth][c]{\includegraphics[]{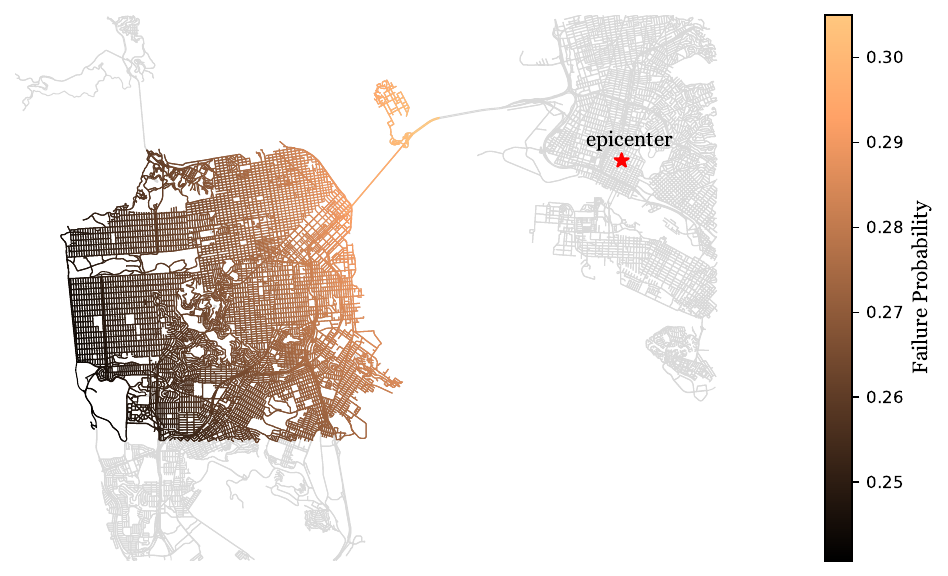}}
    \end{adjustbox}
    \caption{\textbf{Drivable road network in San Francisco.} \textit{The earthquake magnitude is chosen as $10$ for illustration of this figure. The network is very large and computationally infeasible for the Ising model. We employ the dichotomized Gaussian model as a surrogate.}}
    \label{fig:community_network}
\end{figure} 

\subsection{Origin-Destination pairs}
\noindent We construct the hourly Origin-Destination (OD) pairs based on San Francisco County Transportation Authority's ``TNC Today'' report. The TNC data is believed to reflect Uber/Lyft pick-ups and drop-offs in the city by Traffic Analysis Zone (TAZ) \cite{erhardt2022transportation}. Fig.~\ref{fig:OD_matrices} shows the average weekly accumulated OD demands according to the TAZ-level partition. The data can only reflect the OD demands at the coarse-grained TAZ level. As shown in Algorithm~\ref{alg:mat_adj}, we derive the OD matrix through iterative matrix adjustment to make the OD demands consistent with the TAZ level values. In the algorithm, the target $\mathrm{O}_{i}$ or target $\mathrm{D}_{i}$ indicates the Origin or Destination demand at the $i$th TAZ.

\begin{algorithm}
\caption{Matrix adjustment algorithm}
\begin{algorithmic}[1] 
\State Initialize the OD matrix, $\mathrm{OD}_{ij}=1$, $i,j = 1,\dots,n$
\While{$\epsilon > \epsilon_{0}$}\Comment{Determine if the iteration converges}
    \State $\mathrm{OD}_{ij}\leftarrow\mathrm{OD}_{ij}\times \frac{\mathrm{target}\ \mathrm{D}_{j}}{\sum_{i}\mathrm{OD}_{ij}}$ \Comment{Column adjustment}
    \State $\mathrm{OD}_{ij}\leftarrow\mathrm{OD}_{ij}\times \frac{\mathrm{target}\ \mathrm{O}_{i}}{\sum_{j}\mathrm{OD}_{ij}}$ \Comment{Row adjustment}
    \State $\epsilon = \sum_{j}\left|\mathrm{target}\ \mathrm{D}_{j} - \sum_{i}\mathrm{OD}_{ij}\right|$ \Comment{Calculate the error}
\EndWhile
\State \Return $\mathrm{OD}_{ij}$
\end{algorithmic}
\label{alg:mat_adj}
\end{algorithm}

\begin{figure}[H]
    \centering
    \begin{adjustbox}{minipage=\linewidth,scale=0.9}
    \makebox[\textwidth][c]{
    \begin{subfigure}[c]{90mm}
      \centering
      \includegraphics[width=90mm]{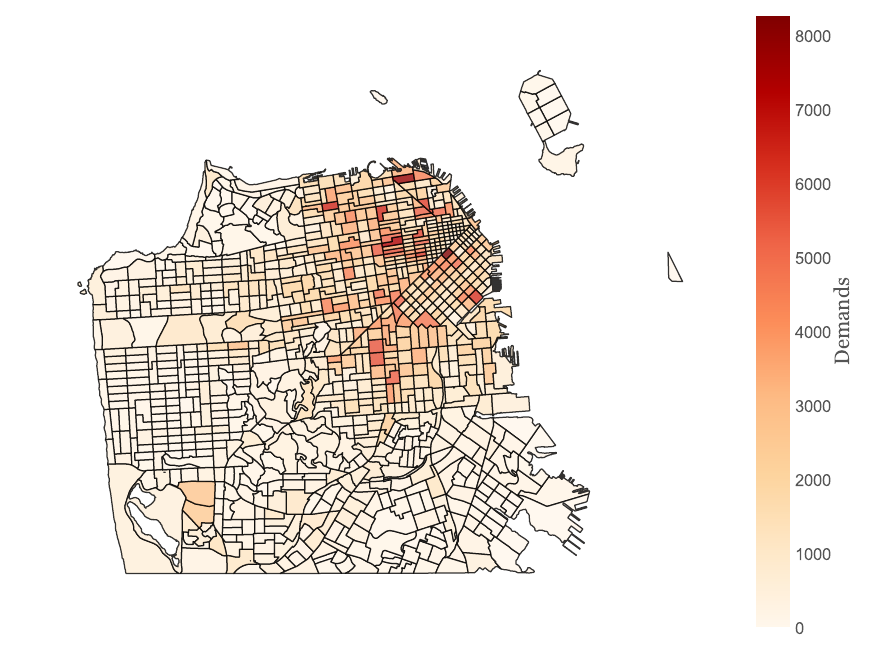}
      \caption{\textbf{Origin points}}
      \label{fig:O}
    \end{subfigure}
    \begin{subfigure}[c]{90mm}
      \centering
      \includegraphics[width=90mm]{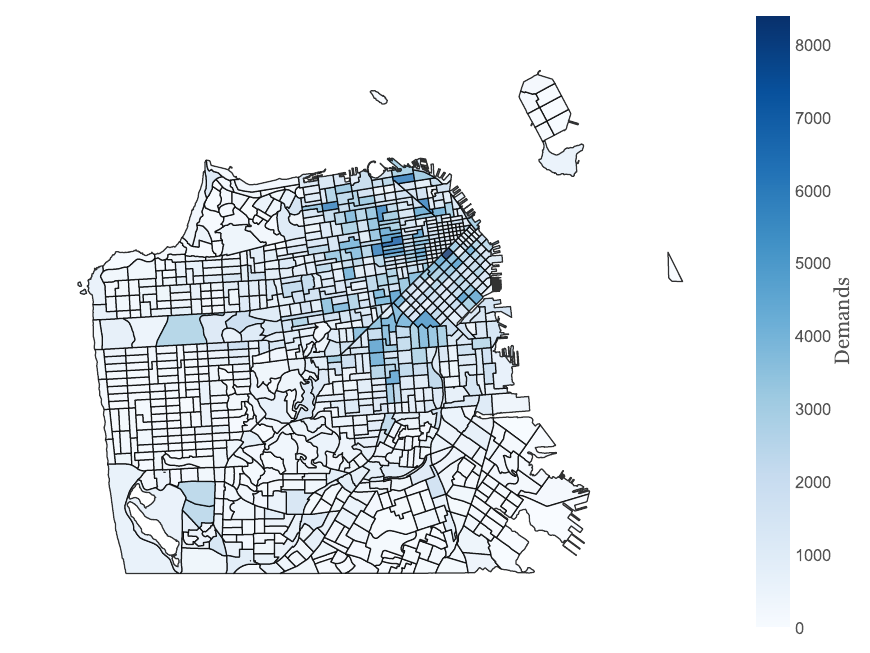}
      \caption{\textbf{Destination points}}
      \label{fig:D}
    \end{subfigure}
    }
    \end{adjustbox}
    \caption{\textbf{Average weekly accumulated OD demands in San Francisco.} \textit{The colorbar represents the Origin/Destination demands in each TAZ.}}
    \label{fig:OD_matrices}
\end{figure}

\subsection{Collective behaviors in trip completion rate}
\noindent We investigate the post-earthquake road network functionality in terms of the trip completion rate given the commuting pattern illustrated by Fig.~\ref{fig:OD_matrices}. Specifically, for each OD pair, we examine if there is at least one route; this event can be represented by a Bernoulli variable $B_i\in\{0,1\}$. The trip completion rate is a random variable defined as $\frac{1}{n_{od}}\sum_iB_i$, where $n_{od}$ is the number of OD pairs. For specific earthquake magnitudes, we investigate the joint distribution of the trip completion rate and the road removal rate to generate  Fig.~\ref{fig:traffic_comp_ratio}. The figure also shows the scenario with no correlation in road failures, highlighted by the red circles. The motivation of Fig.~\ref{fig:traffic_comp_ratio} is to project topology (road removal rate), functionality (trip completion rate), and hazard intensity (earthquake magnitude) into the same figure to investigate the collective behaviors of the topological and functional changes of a road network influenced by earthquakes. It is observed that the product outcome space of road removal rate and trip completion rate exhibits two phases and the increase of earthquake magnitude stimulates the transition from one phase to another; while if there is no correlation in road failures, there is only one phase. It follows that the tendency of road components to ``fail" or ``work" collectively serves as a double-edged sword: for relatively small hazard intensity, there is a mild chance for system-level malfunction; for relatively large hazard intensity, there is also a mild chance of global functioning. Finally, it is worth mentioning that Fig.~\ref{fig:OD_matrices} shares some features of a percolation analysis \cite{zeng2019switch,li2015network,di2023entropy,zhong2020network,behrensdorf2021numerically}, but it offers more information and diverges from percolation in many aspects--the road removal rate in Fig.~\ref{fig:OD_matrices} is not a ``free" or control parameter, and the goal is to discover probabilistic patterns in functionality and topology, rather than to identify the critical road removal rate to trigger a phase transition or the distribution of functionality metrics at the criticality.

\begin{figure}[H]
    \centering
    \begin{adjustbox}{minipage=\linewidth,scale=0.87}
    \makebox[\textwidth][c]{\includegraphics[]{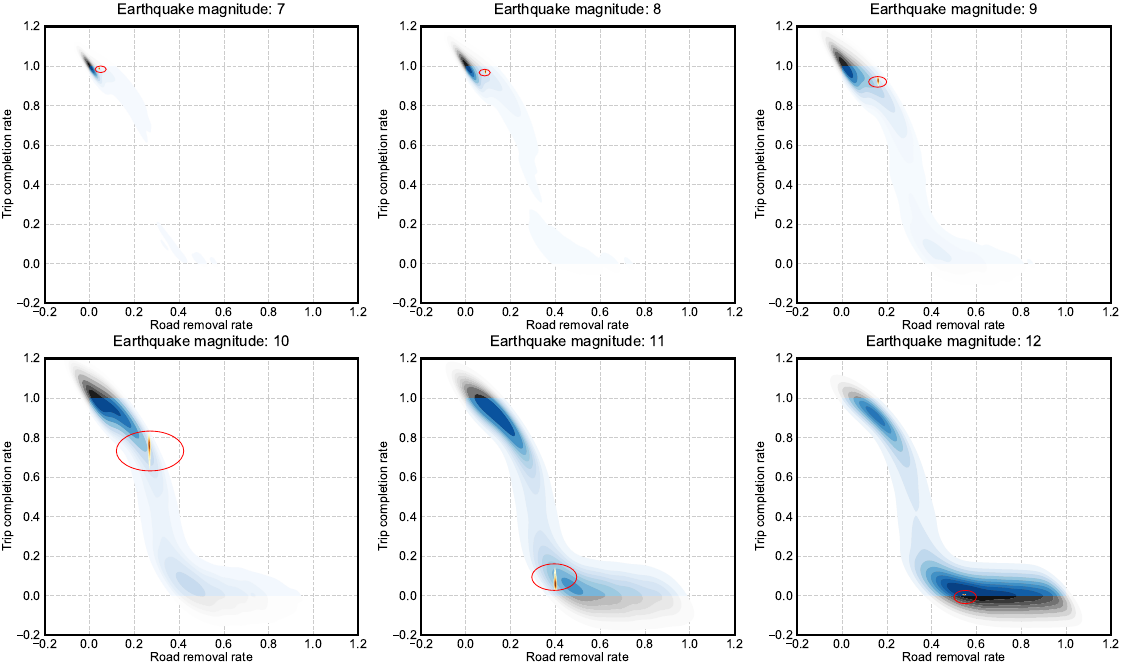}}
    \end{adjustbox}
    \caption{\textbf{Joint distributions of post-earthquake trip completion rate and road removal ratio under increasing earthquake magnitudes}. \textit{This figure shows the joint distributions of trip completion rate and road removal ratio influenced by earthquake magnitudes. For a comparison, the distributions with no correlations in road failures are depicted in red and highlighted by the red circles. It is observed that the product outcome space of road removal rate and trip completion rate consistently exhibits two phases regardless of the earthquake magnitude--the magnitude alters the weights of the two phases and an increase of magnitude stimulates the transition from one phase to another. If there is no correlation in road failures, there is only one phase--an outcome anticipated by the Central Limit Theorem. The figure also suggests a percolation phase transition: with the increase in road removal ratio, the drop in traffic completion rate can be abrupt in a critical interval. We used kernel density estimation to generate this figure, yielding density contours beyond the feasible region of $[0,1]^2$, which are marked by gray}. }
    \label{fig:traffic_comp_ratio}
\end{figure}

\section{Additional remarks}\label{Sec:Addi}
\noindent The maximum entropy model can be extended to involve temporal evolution \cite{ding2009mixing,heyl2013dynamical}, such as the Ising model with Glauber dynamics \cite{glauber1963time}. A time-dependent maximum entropy model can be utilized to study how a complex infrastructure evolves and recovers after a natural hazard. Furthermore, the coupling effects are not necessarily symmetric \cite{crisanti1988dynamics,ginzburg1994theory,aguilera2023nonequilibrium,merle2019turing}. This asymmetric coupling becomes important if we model functional dependencies between infrastructures. For example, a hospital may rely on a power transmission infrastructure to operate, but not vice versa. Therefore, it is promising to extend the maximum entropy modeling framework to investigate the time-dependent collective behaviors of civil infrastructure systems with complex asymmetric functional dependencies. 

\section{Conclusions}\label{Sec:Concl}
\noindent In this study, we present maximum entropy modeling for the regional hazard responses of civil infrastructure systems. For the special but typical case where the performance states of individual structures are binary, and the mean performance state values and their pairwise correlations are given, the maximum entropy model reduces to the Ising model in statistical physics. For this special case, we propose using a dichotomized Gaussian distribution as a near-maximum entropy surrogate model. This surrogate model has satisfactory scalability, which is then applied to study the post-earthquake functionality of the drivable road network in San Francisco, consisting of $8,694$ nodes and $26,964$ links. We have observed that the joint outcome space of road removal ratio and trip completion rate exhibits two patterns, and the increase of earthquake magnitude stimulates the transition from one to another; while if there is no correlation in road failures, there is only one trivial pattern induced by the Central Limit Theorem. This paper focuses on the probabilistic modeling of hazard responses of infrastructure systems, which is the foundation for understanding a community's post-hazard behaviors and resilience.} For future research, we will adapt and adopt the maximum entropy model for various natural hazards and investigate the temporal evolution of the collective behaviors of infrastructure systems.

\section*{Code and data acquisition}
{All codes and data are made available upon reasonable requests.}
\section*{CRediT authorship contribution statement}

{\textbf{Xiaolei Chu}}: Writing - original draft, Conceptualization, Formal analysis, Investigation, Methodology, Validation, Visualization.
{\textbf{Ziqi Wang}}: Conceptualization, Supervision, Writing - review $\&$ editing, Funding acquisition.

\section*{Acknowledgments}
We thank Dr.~Jianhua Xian for the help in estimating the entropy using a variance reduction method. Thanks also go to Dr.~Dongkyu Lee for providing the seismic hazard model.

\section*{Declaration of competing interest}
The authors declare that they have no known competing financial interests or personal relationships that could have appeared to influence the work reported in this paper.

\newpage
\bibliography{Ising_based_surrogate.bib}

\newpage
\appendix
\section{Entropy estimation}
\label{sec:appendixA}
\subsection{The Ising model}
\noindent Recall that the PMF of the Ising model is defined as $p\left(\vect{x};\vect{J}\right)=\exp \left( \vect{x}\tr\vect{J}\vect{x} \right)/Z\left(\vect{J}\right)$. Typically, we can estimate an expectation, such as $\mathbb{E}_{p\left(\vect{x};\vect{J}\right)}\left[f\left(\vect{X}\right)\right]$, without knowing the partition function $Z\left(\vect{J}\right)$, but the entropy explicitly involves  the partition function. Specifically, the entropy for the Ising model is
\begin{equation}
    \mathbb{E}_{p\left(\vect{x};\vect{J}\right)}\left[ -\log p\left(\vect{X};\vect{J}\right) \right] = \mathbb{E}_{p\left(\vect{x};\vect{J}\right)}\left[ -\vect{X}\tr \vect{J} \vect{X} \right] + Z\left(\vect{J}\right)\,,
\end{equation}
where the evaluation of the partition function is the computational bottleneck. We adapt the sequential Monte Carlo \cite{del2006sequential,xian2024relaxation} to this problem by defining $Z_{n} = \int_{\Omega_{\vect{x}}} \exp{\left(\vect{x}\tr\vect{J}\vect{x}\right/T_{n})}\mathrm{d}\vect{x}$, where $T_{n}$ is the ``temperature,`` used as an annealing parameter at the $n$th iteration. It follows that $Z_n$ can be rewriten as 
\begin{equation}
\begin{aligned}
    Z_{n} & = \sum_{\vect x} \exp{\left(\vect{x}\tr\vect{J}\vect{x}/T_{n}\right)}\\
    & = \sum_{\vect x}\frac{\exp{\left( \vect{x}\tr\vect{J}\vect{x}/T_{n} \right)}}{\exp{\left(\vect{x}\tr\vect{J}\vect{x}\right/T_{n-1})}/Z_{n-1}} \frac{\exp{\left(\vect{x}\tr\vect{J}\vect{x}\right/T_{n-1})}}{Z_{n-1}} \,.
\end{aligned}
\end{equation}
Then we have
\begin{equation}
\begin{aligned}
    \frac{Z_{n}}{Z_{n-1}} & = \sum_{\vect x} \frac{\exp{\left( \vect{x}\tr\vect{J}\vect{x}/T_{n} \right)}}{\exp{\left(\vect{x}\tr\vect{J}\vect{x}\right/T_{n-1})}} \frac{\exp{\left(\vect{x}\tr\vect{J}\vect{x}\right/T_{n-1})}}{Z_{n-1}} \\
    & = \mathbb{E}_{p\left(\vect{x};\vect{J}/T_{n-1}\right)}\left[ \frac{\exp{\left( \vect{x}\tr\vect{J}\vect{x}/T_{n} \right)}}{\exp{\left(\vect{x}\tr\vect{J}\vect{x}\right/T_{n-1})}} \right]\,.
\end{aligned}
\end{equation}
We start from a high temperature $T_{1}$, such that the distribution is dispersed, and stop at $T_{N_{T}}=1$ to restore the original partition function. We choose $T_{n}=1.6^{\frac{20}{N_{T}}\left(N_{T}-n\right)}$ with $N_{T}=100$ to tune the annealing process, and the partition function is estimated as
\begin{equation}
    Z\left(\vect{J}\right) = Z_{N_{T}} \approx Z_{0} \prod_{n=1}^{N_{T}} \frac{Z_{n}}{Z_{n-1}}\,,
\end{equation}
where $Z_{0}$ is chosen as the partition function of a multivariate independent Bernoulli distribution with marginals $p_{i}\left(1\right)=p_{i}\left(0\right)=0.5$ and $Z_{0}=2^{d}$. $d$ is the dimension of $\vect{x}$.

Finally, the entropy is estimated by
\begin{equation}
\begin{aligned}
    \mathbb{E}_{p\left(\vect{x};\vect{J}\right)}\left[ -\log p\left(\vect{X};\vect{J}\right) \right] & = \sum_{\vect x}\left(-\vect{x}\tr\vect{J}\vect{x}\right) p\left(\vect{x};\vect{J}\right) + Z\left(\vect{J}\right) \\
    & \approx -\sum_{i=1}^{N_{0}} \vect{x}_{i}\tr\vect{J}\vect{x}_{i} + Z_{0} \prod_{n=1}^{N_{T}} \frac{Z_{n}}{Z_{n-1}}
\end{aligned}
\end{equation}
where $\vect{x}_{i}$ are samples from the Ising model and $N_{0}=10^{5}$ is used in this study.

\subsection{Dichotomized Gaussian model}
\noindent Recall that the PMF of the dichotomized Gaussian distribution is
\begin{equation}
\label{eq:dg}
q(\vect{x})=\frac{1}{(2 \pi)^{d / 2}|\vect{\Lambda}|^{1 / 2}} \int_{a(x_1)}^{b(x_1)} \cdots \int_{a(x_d)}^{b(x_d)} \exp \left(-\frac{1}{2}(\vect{z}-\vect{\gamma})\tr \vect{\Lambda}^{-1}(\vect{z}-\vect{\gamma})\right)\,\mathrm{d}\vect{z}\,,
\end{equation}
where $a_{i}=0$, $b_{i}=\infty$ if $x_{i}=1$; $a_{i}=-\infty$, $b_{i}=0$ if $x_{i}=0$. In this case, we need to use Monte Carlo simulation to estimate both $q\left(\vect{x}\right)$ and the entropy $\mathbb{E}_{q\left(\vect{x}\right)}\left[ -\log q\left(\vect{X}\right) \right]$. The samples to estimate $q\left(\vect{x}\right)$ are generated by the Gaussian distribution $\mathcal{N}\left(\vect{\gamma},\vect{\Lambda}\right)$, while the samples to estimate $\mathbb{E}_{q\left(\vect{x}\right)}\left[ -\log q\left(\vect{X}\right) \right]$ are from Eq.~\eqref{eq:dicho} and Eq.~\eqref{eq:paraest}. For $q\left(\vect{x}\right)$, we have
\begin{equation}
\label{eq:est_qx}
    q\left(\vect{x}\right)=\int_{\vect{z} \in \Omega_{\vect{x}}} f\left(\vect{z};\vect{\gamma},\vect{\Lambda}\right)\mathrm{d}\vect{z}=\int_{\vect{z}\in \mathbb{R}^{d}} \mathds{1}\left(\vect{z} \in \Omega_{\vect{x}}\right) f\left(\vect{z};\vect{\gamma},\vect{\Lambda}\right) \mathrm{d}\vect{z}\,,
\end{equation}
where $\Omega_{\vect{x}}=\prod_{i=1}^{d}\left[a\left(x_{i}\right),b\left(x_{i}\right)\right]$ is the integral domain; $\mathds{1}\left(\cdot\right)$ is the indicator function;  $f\left(\vect{y};\vect{\gamma},\vect{\Lambda}\right)$ is the Gaussian density in Eq.~\eqref{eq:dg}. Eq.~\eqref{eq:est_qx} is in the classic form of reliability analysis and rare-event simulation, so the estimation of $q\left(\vect{x}\right)$ can be handled by rare event simulation techniques \cite{xian2024relaxation,au2001estimation,wang2019hamiltonian,chen2022riemannian}. We do not repeat the details here. The estimation of the entropy $\mathbb{E}_{q\left(\vect{x}\right)}\left[ -\log q\left(\vect{X}\right) \right]$ is performed by direct Monte Carlo simulation. 

\end{document}